\documentstyle[aps,prl]{revtex} 
\begin{document} 
\twocolumn 
\input epsf 
\newcommand{\epo}{\epsilon(\omega)} 
\newcommand{\enul}{\epsilon_0} 
\newcommand{\de}{\Delta\epsilon} 
\newcommand{\sio}{\sigma(\omega)} 
\newcommand{\sitm}{\tilde{\sigma}'} 
\newcommand{\siDC}{\sigma(0)} 
\newcommand{\sit}{\tilde{\sigma}} 
\newcommand{\omt}{\tilde{\omega}} 
\newcommand{\stg}{s\tilde{G}} 
 
\title{Scaling and universality of AC conduction in disordered solids} 
 
\author{Thomas B. Schr{\o}der\cite{tbs} and Jeppe C. Dyre} 
 
\address{Department of Mathematics and Physics (IMFUFA), Postbox 
260, Roskilde University, DK-4000 Roskilde, Denmark} 
 
\maketitle

\begin{abstract} 
Recent scaling results for the AC conductivity of ionic glasses 
by Roling {\it et al.} [Phys. Rev. Lett. {\bf 78}, 2160 (1997)] 
and Sidebottom [Phys. Rev. Lett. {\bf 82}, 3653 (1999)] are 
discussed. It is shown that Sidebottom's version of scaling is 
completely general. A new analytical approximation to the 
universal AC conductivity of hopping in the extreme disorder 
limit, the ``diffusion cluster approximation,'' is presented and 
compared to simulations and experiments. 
\end{abstract}

 
Disordered solids have AC electrical properties remarkably in common. 
These solids, in fact, have so similar frequency-dependent conductivity 
$\sio$ that ionic conduction cannot be distinguished from electronic. Even 
the temperature-dependence of $\sio$ is ``quasi-universal.'' 
 
The class of disordered solids with quasi-universal AC behavior 
is large, including polycrystalline and amorphous semiconductors, 
organic semiconductors, ionic conductive glasses, ionic melts, 
non-stoichiometric crystals, ionic or electronically conducting 
polymers, metal cluster compounds, transition metal oxides, and 
macroscopic mixtures of differently conducting phases like 
organic-inorganic composites, etc. Each class contains hundreds 
of different solids and there is a huge literature on their AC 
conductivities. 
 
It is possible for almost all disordered solids to scale AC data 
at different temperatures into one single curve. This 
so-called master curve gives the dimensionless AC conductivity 
$\sit\equiv\sio/\siDC$ as function of a scaled dimensionless 
frequency $\omt$. The existence of a master curve is sometimes 
referred to as the ``time-temperature superposition principle'' 
(TTSP). The master curves of different solids - while not 
identical - are surprisingly similar. This quasi-universality was 
recognized gradually in the 1970's 
\cite{isa70,jon77,owe77,man80}. 
 
The common AC features of disordered solids are the following 
\cite{dyr88}: At low frequencies conductivity is 
frequency-independent. Around the dielectric loss peak frequency 
$\omega_m$ \cite{diel_def} AC conduction sets in, and for 
$\omega\gg\omega_m$ $\sio$ is close to a frequency power-law. The 
exponent is between 0.7 and 1.0 \cite{0.6}. As temperature is 
lowered the exponent goes to 1.0. In a log-log plot the AC 
conductivity is much less temperature-dependent than the DC 
conductivity. A final ubiquitous observation is the 
Barton-Nakajima-Namikawa (BNN) relation 
\cite{bar66,nak72,nam75,tom78} connecting dielectric loss 
strength $\de$\cite{diel_def}, dielectric loss peak frequency, 
and DC conductivity - $\siDC\ =\ p\ \de\ \enul\ \omega_m$ - where 
$p$ is a numerical constant close to one. 
 
To construct the master-curve frequency must be scaled by 
$\omega_m$. Because the dielectric loss strength is only weakly 
temperature-dependent while $\siDC$ and $\omega_m$ are both 
Arrhenius, the BNN-relation implies $\omega_m\sim\siDC$. Thus, 
the existence of a master curve is conveniently summarized into 
\begin{equation}\label{2} 
\sit =\ F\left(\frac{C}{\siDC}\ \omega\right)\ \equiv\ F(\omt)\,, 
\end{equation} 
where $C$ may depend on variables like charge carrier 
concentration $n$, temperature $T$, high frequency dielectric 
constant, etc. 
 
Recently, there has been renewed interest in scaling and 
universality of AC data for ionic glasses 
\cite{rol97,fun97,rol98,sid99} (excellent reviews of glass ionic 
conduction have been given by Nowick and coworkers \cite{now94} 
and Ngai \cite{nga96}). In 1997 Roling, Happe, Funke, and Ingram 
showed that $C\propto n/T$ for sodium borate glasses 
\cite{rol97}. This year, however, Sidebottom showed that in general 
scaling is not achieved by  $C\propto n/T$ - instead three quite 
different ionic conductive systems all obey the following scaling relation
\cite{sid99}, 
 
\begin{equation}\label{3} 
\sit =\ F\left(\frac{\enul\de}{\siDC}\ \omega\right)\ \equiv\ F(\omt)\,. 
\end{equation} 
 
The purpose of this paper is three-fold. First, we put the recent 
scaling results into a historical perspective. Second, we show 
that Eq.\ (\ref{3}) is generally valid. Finally, Sidebottom's 
experimental master curves are compared to two models, the 
symmetric hopping model and the macroscopic model. In this connection a new 
analytical approximation to the universal AC hopping conductivity 
is presented. 
 
First, the historical perspective. In a series of papers towards 
the end of the 1950's Taylor analyzed the dielectric properties 
of ionic glasses in accordance with the Debye equation with a 
spread of relaxation times \cite{tay}. He showed that the 
dielectric loss \cite{diel_def} for all glasses fell on a single 
plot against scaled frequency. In 1962 Isard relabeled Taylor's 
axis by plotting dielectric loss against log of the product of 
frequency and resistivity and thus essentially arrived at AC 
scaling in the form given in Eq.\ (\ref{2}) \cite{isa62}. Since 
then Eq.\ (\ref{2}), which we shall refer to as ``Taylor-Isard 
scaling,'' has been used in several different contexts. For 
instance, Taylor-Isard scaling was used by Scher and Lax in 1973 
in their famous papers introducing the continuous time random 
walk approximation \cite{sch73}, by Summerfield and coworkers in 
1985 for amorphous semiconductors \cite{sum85,bal85}, by van 
Staveren and coworkers in 1991 for metal-cluster compounds 
\cite{van91}, and by Kahnt the same year for ionic glasses 
\cite{kah91}. 
 
We now proceed to comment on the scaling principle of Roling and 
coworkers, $C\propto n/T$ in Eq.\ (\ref{2}), and Sidebottom's 
recent interpretation of its occasional violations. If $q$ is 
charge, $f_H$ jump rate, and $d$ jump length, Sidebottom bases 
his arguments on the expressions $\siDC\propto nq^2d^2 f_H/k_BT$ 
and $\de\propto nq^2d^2/k_BT$ \cite{sid99}. Clearly, to obtain 
$\omt$ frequency should be scaled by $f_H$, thus leading to 
$C\propto n/T$ if all other parameters are constant. But since it 
is reasonable to assume the jump length would increase as 
concentration decreases, Sidebottom argues that one cannot expect 
$C\propto n$; on the other hand the $n$-dependence of $d$ is 
taken care of by Eq.\ (\ref{3}) in which the unknown jump length 
is eliminated by scaling with the measured $\de$. 
 
In our opinion Sidebottom's arguments are largely correct, but 
Eq.\ (\ref{3}) is much more general than it appears from his 
Letter. Consider hopping of completely non-interacting charge 
carriers on a cubic lattice. In this model conductivity is $n$ 
times charge carrier mobility and for given lattice jump 
frequencies the mobility scales with $d^2$. This shows that 
$C\propto nd^2$ as Sidebottom has it. However, if the lattice is 
homogeneous with just one jump frequency conductivity is 
frequency-independent; in order to have strongly 
frequency-dependent conductivity lattice jump frequencies must 
vary many decades. In such more realistic cases the expressions 
used by Sidebottom for $\siDC$ and $\de$ do not apply - there 
simply is no unique $f_H$. Still, we find in our simulations that 
these more realistic hopping models always obey Eq.\ (\ref{3}). 
This leads to the question: When is Eq.\ (\ref{3}) obeyed? 
 
Equation (\ref{3}) apparently expresses {\em two} informations: 
1) TTSP is obeyed, and 2) the scaled frequency is given by 
$\omt=[\enul\de/\siDC]\ \omega$. We now show that 1) 
mathematically implies 2): TTSP implies the existence of some 
function $\sit(\omt)$ where $\omt$ is the scaled frequency. 
Expanding to first order in $\omt$ leads to $\sit=1+iA\omt$ 
(where $A$ is real because $\sigma^*(\omega)=\sigma(-\omega)$). 
Since $\sigma=\sit\siDC$ we have $\sigma=\siDC+iA\omt\siDC$. On 
the other hand, from the definition of $\de$ \cite{diel_def} one 
has $\sigma=\siDC+i\omega\de\enul$ for $\omega\rightarrow 0$. 
Equating the two expressions for $\sigma$ leads to $\omt=A^{- 
1}[\de\enul/\siDC]\ \omega$, i.e., Eq.\ (\ref{3}) 
\cite{anal_note} (the numerical value of $A$ is irrelevant). 
Thus, Eq.\ (\ref{3}) follows from TTSP alone. 
 
We next compare the data discussed by Sidebottom to two models, 
the macroscopic model and the symmetric hopping model. The macroscopic 
model considers AC conduction in a random mixture of phases with 
different (frequency-independent) conductivities 
\cite{fis86,dyr93}. The AC conductivity is obtained by solving 
Maxwell's equations. Computer simulations have shown 
\cite{dyr93,rie94} that in the {\em extreme disorder limit} 
(where the local conductivities vary over many decades) the AC conductivity 
is accurately described by the effective medium approximation 
(EMA) universality equation \cite{fis86,dyr93,bry80}, 
 
\begin{equation}\label{4} 
\sit\ln\sit\ =\ i\omt\,. 
\end{equation} 
The symmetric hopping model \cite{bot85,hau87,dyr94}, which models AC 
conduction via random walks of non-interacting particles on a 
cubic lattice with random symmetric Arrhenius nearest-neighbor jump rates 
($\gamma_0 e^{-\beta E}$, where $E$ is the random energy barrier), 
also exhibits universality in the extreme disorder limit. Even for this
model does the EMA lead to Eq.\ (\ref{4}) \cite{dyr94}. For hopping, 
however, Eq.\ (\ref{4}) does not give an accurate representation 
of the universal $\sit(\omt)$ \cite{dyr94,dyr96,sch99}. 
 
\begin{figure} 
  \hbox to\hsize{\epsfxsize=.95\hsize\hfil\epsfbox{ 
  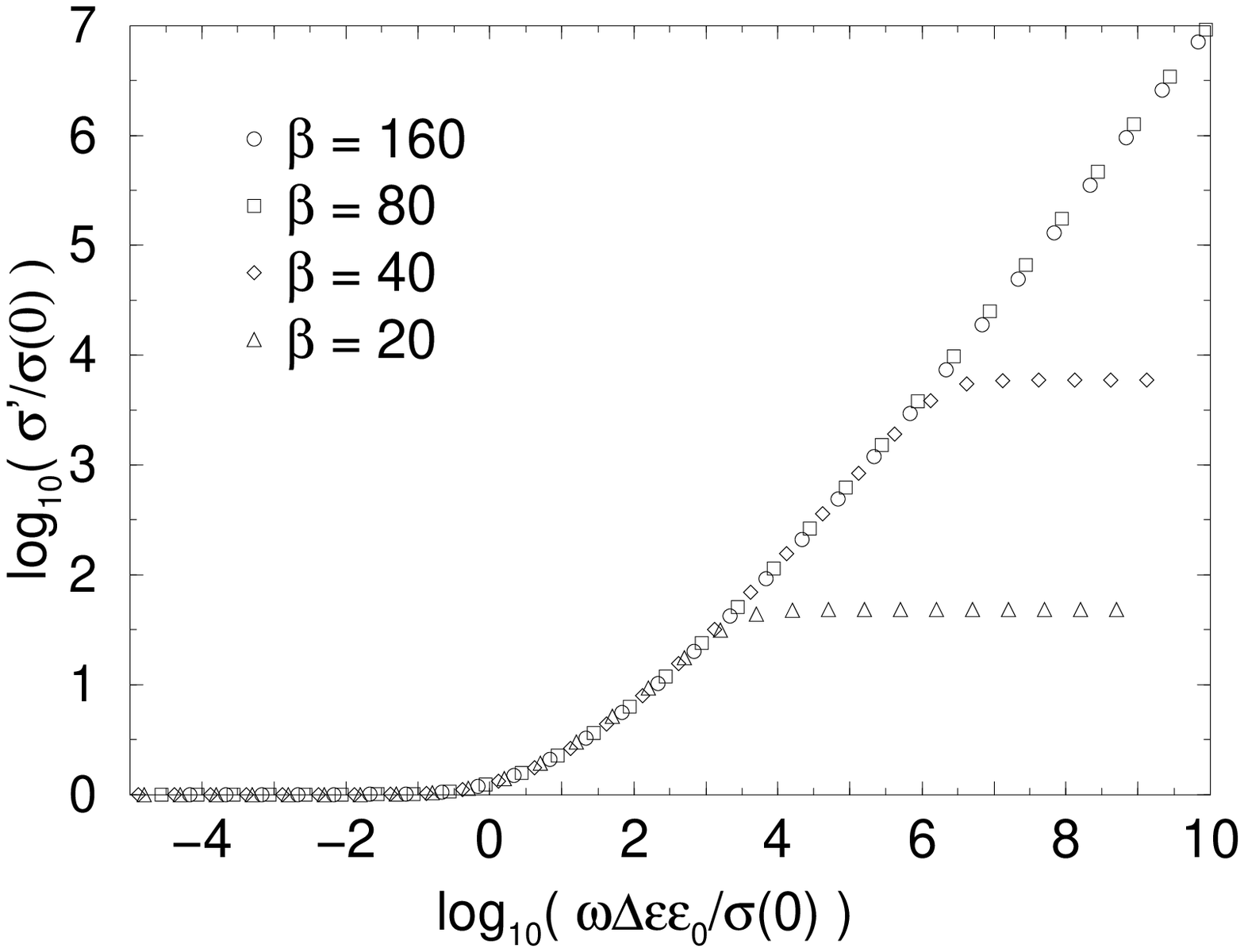}\hfil} 
  \hbox to\hsize{\epsfxsize=.95\hsize\hfil\epsfbox{ 
  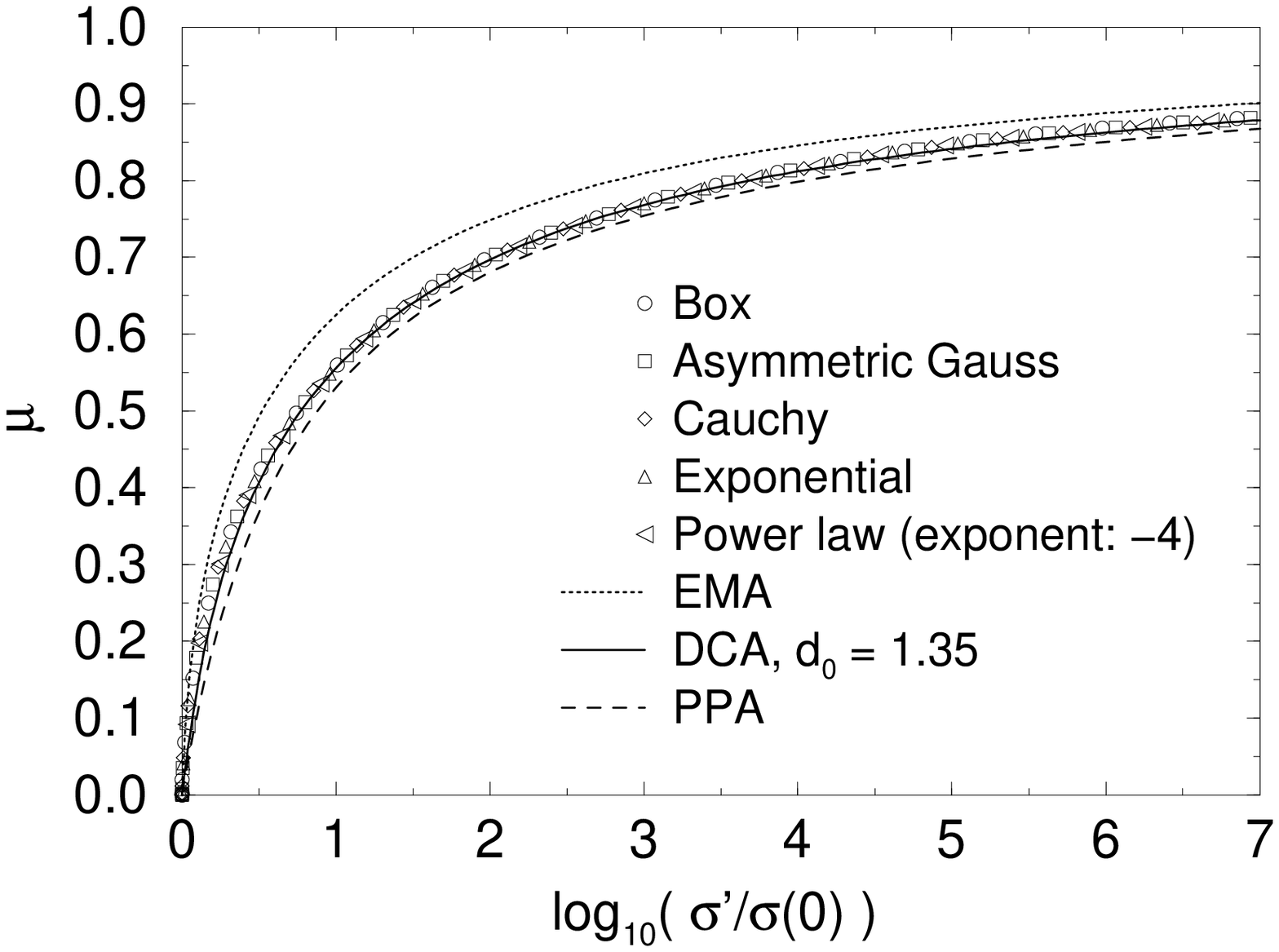}\hfil} 
\caption{ 
Numerical results for the symmetric hopping model in 3 dimensions with 
periodic boundary conditions \protect\cite{sch99}. Reported results are averages 
over 100 independent $N\times N\times N$ cubic lattices. 
a) Real part of $\sigma (\omega)$ for the Box distribution of energy 
barriers ($p(E)=1,~ 0\le E \le 1$), 
scaled according to Eq.\ (\ref{3}); 
$\beta=20$ ($N=14$), $\beta=40$ ($N=24$), $\beta=80$ ($N=32$), 
and  $\beta=160$ ($N=64$). As $\beta$ increases the data converge 
to a universal curve. 
b) The apparent exponent $\mu\equiv d\ln\sitm/d\ln\omt$ plotted 
versus $\sitm$. Data are shown for 5 different energy barrier distributions,
$p(E)$, see \protect\cite{dyr93} for details.  $\tilde \beta = 160$  ($N=64$), where
$\tilde \beta \equiv \beta/p(E_c)$ and $E_c$ is the ``percolation 
energy'' \protect\cite{dyr93,dyr96}. The universal curve is independent 
of the energy barrier distribution. Estimated values of $\de\enul/\beta$ 
for different distributions agree within 1\%. 
The numerical data are compared 
to three analytical approximations: EMA (Eq.\ (\ref{4})), PPA (\protect\cite{ppa}), 
and DCA (Eq.\ {\ref{5}}). The universal curve lies between EMA and PPA 
and is well approximated by DCA with $d_0=1.35$. 
} 
\end{figure} 
 
Before comparing model predictions to experimental data we present results 
from  new simulations of the symmetric hopping model. The simulations 
were done on 3-dimensional samples with periodic boundary conditions, 
using a method based on exactly solving the master equation \cite{sch99}. 
Figure 1a shows $\sitm$ - the real part of $\sit$ -  for the Box 
distribution of energy barriers. The frequency axis is scaled according to 
Eq.\ (\ref{3}). Clearly, as $\beta\equiv 1/k_BT$ increases $\sitm$ converges
to a single curve. As seen in Fig. 1b showing the apparent exponent 
$\mu$ ($\sitm \sim \omt^{\mu}$) as function of $\sitm$, the universal curve 
is the same for different energy barrier distributions. Figure 1b compares 
the simulations to three analytical approximations, the EMA universality 
equation Eq.\ (\ref{4}), the percolation path approximation (PPA) 
\cite{dyr96,ppa}, and a new ``diffusion cluster approximation'' 
(DCA), which we  describe now: 
 
The reason for AC universality in hopping is the fact that in the 
extreme disorder limit conduction takes place on the 
``percolation cluster'' formed by the links with largest jump 
rates until percolation \cite{dyr93,dyr94,dyr99}; the PPA-idea 
\cite{dyr93,dyr96} is to regard the conducting paths on the 
percolation cluster as strictly one-dimensional. The physical 
idea behind EMA is to replace the inhomogeneous lattice by an 
effective homogeneous medium determined self-consistently 
\cite{bot85,hau87}. Our hopping simulations show that the truth 
is somewhere between PPA and EMA \cite{sch99}, somewhere between 
one-dimensionality and homogeneity. Most likely, this is because 
both approximations ignore the fact that conduction takes place 
on some complex subset of the percolation cluster. This 
``diffusion cluster'' must be smaller than the backbone (defined 
by removing dead-ends of the percolation cluster, fractal dimension=1.7 
\cite{sch99,sta92}) and larger than the set of red bonds 
(those that, when cut, stops the current, fractal dimension=1.1 
\cite{sch99,sta92}). At present this is all we know about the diffusion 
cluster and its dimension $d_0$ is regarded below as a fitting parameter. 
To derive the DCA equation we use EMA in $d_0$ dimensions. In the 
extreme disorder limit EMA implies \cite{dyr94,ema_units} 
$\ln\sit\propto\stg$ where, if $p({\bf k})=\frac{1}{d_0} 
\sum_{i=1}^{d_0}\cos (k_i)$, $\stg$ is the following integral 
$\stg=\int_{-\pi<k_i<\pi}\frac{d {\bf 
k}}{(2\pi)^d}\frac{i\omega}{i\omega+2d_0\sigma[1-p(\bf{k})]}$. 
Whenever $1<d_0<2$ one finds $\stg\propto 
(i\omega/\sigma)^{d_0/2}$ at relevant \cite{dyr94} frequencies. 
Thus, after rescaling frequency we arrive at the DCA equation, 
 
\begin{equation}\label{5} 
\ln\sit\ =\ \left(\frac{i\omt}{\sit}\right)^{d_0/2}\,. 
\end{equation} 
As is clear from Fig. 1b, the solution of this equation for 
$d_0=1.35$ gives an excellent fit to the universal AC hopping 
conductivity. 
 
\begin{figure} 
  \hbox to\hsize{\epsfxsize=1.\hsize\hfil\epsfbox{ 
   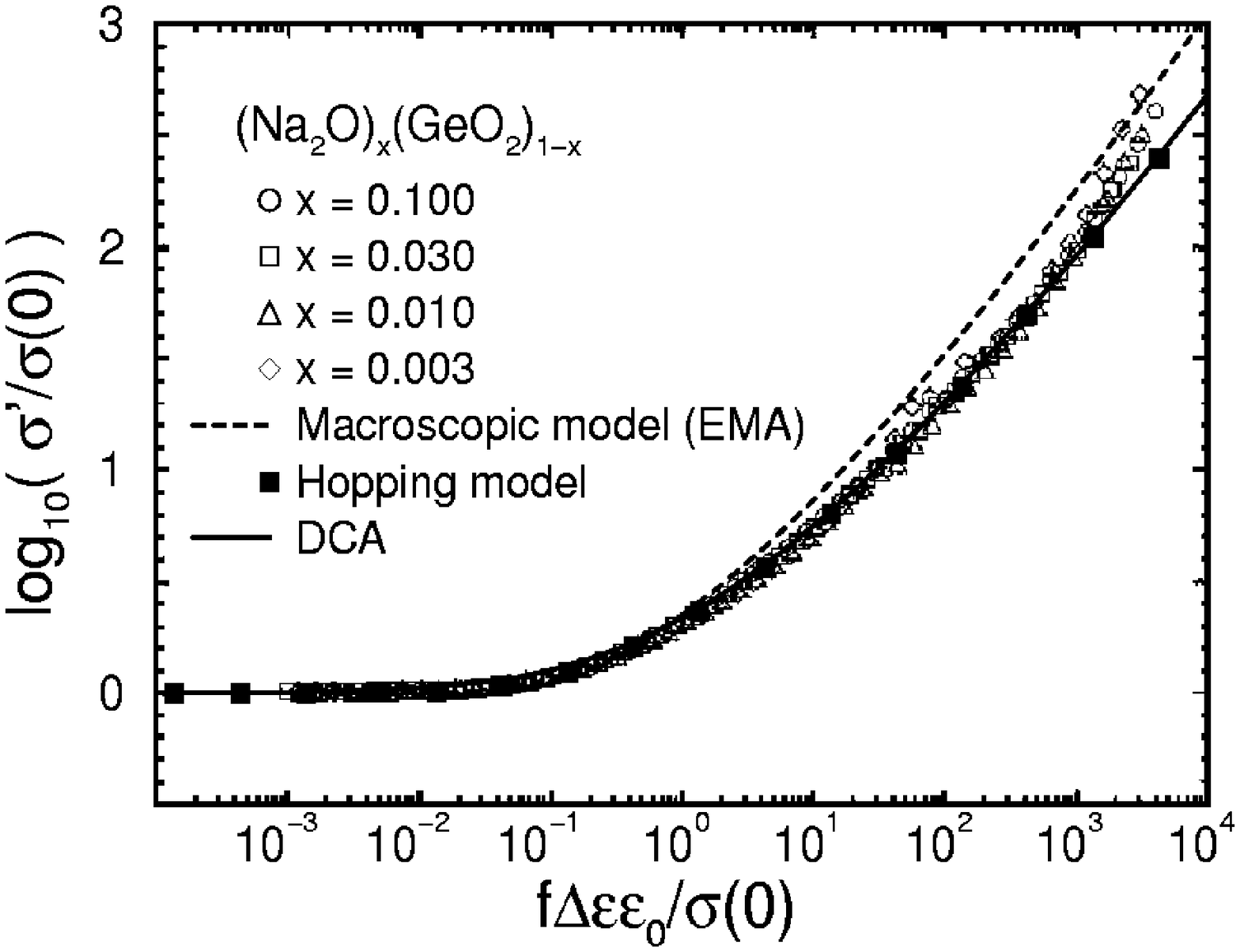}\hfil} 
  \vspace{-.1cm} 
  \hbox to\hsize{\epsfxsize=1.\hsize\hfil\epsfbox{ 
   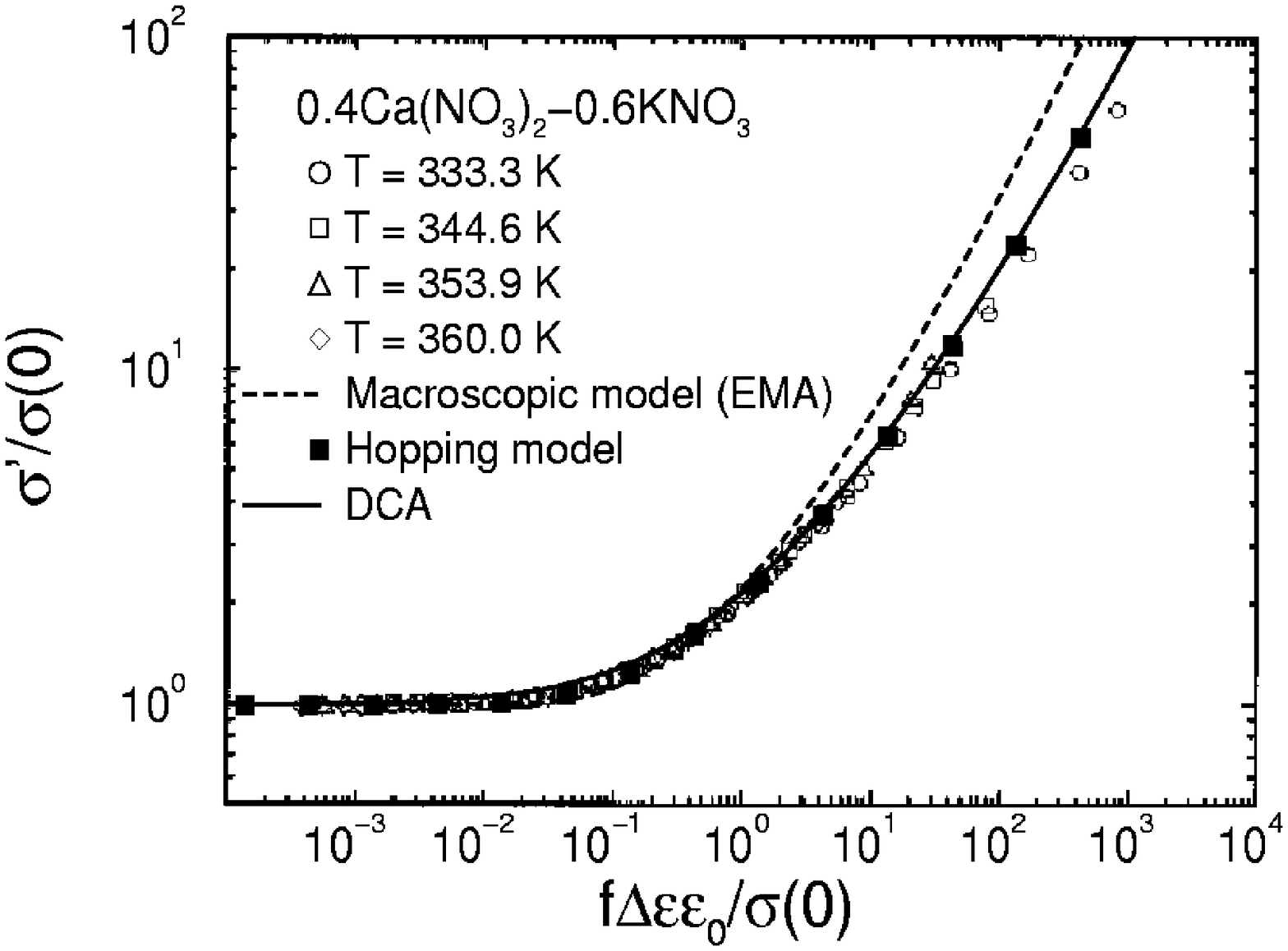}\hfil} 
   \vspace{-.1cm} 
   \hbox to\hsize{\epsfxsize=1.\hsize\hfil\epsfbox{ 
   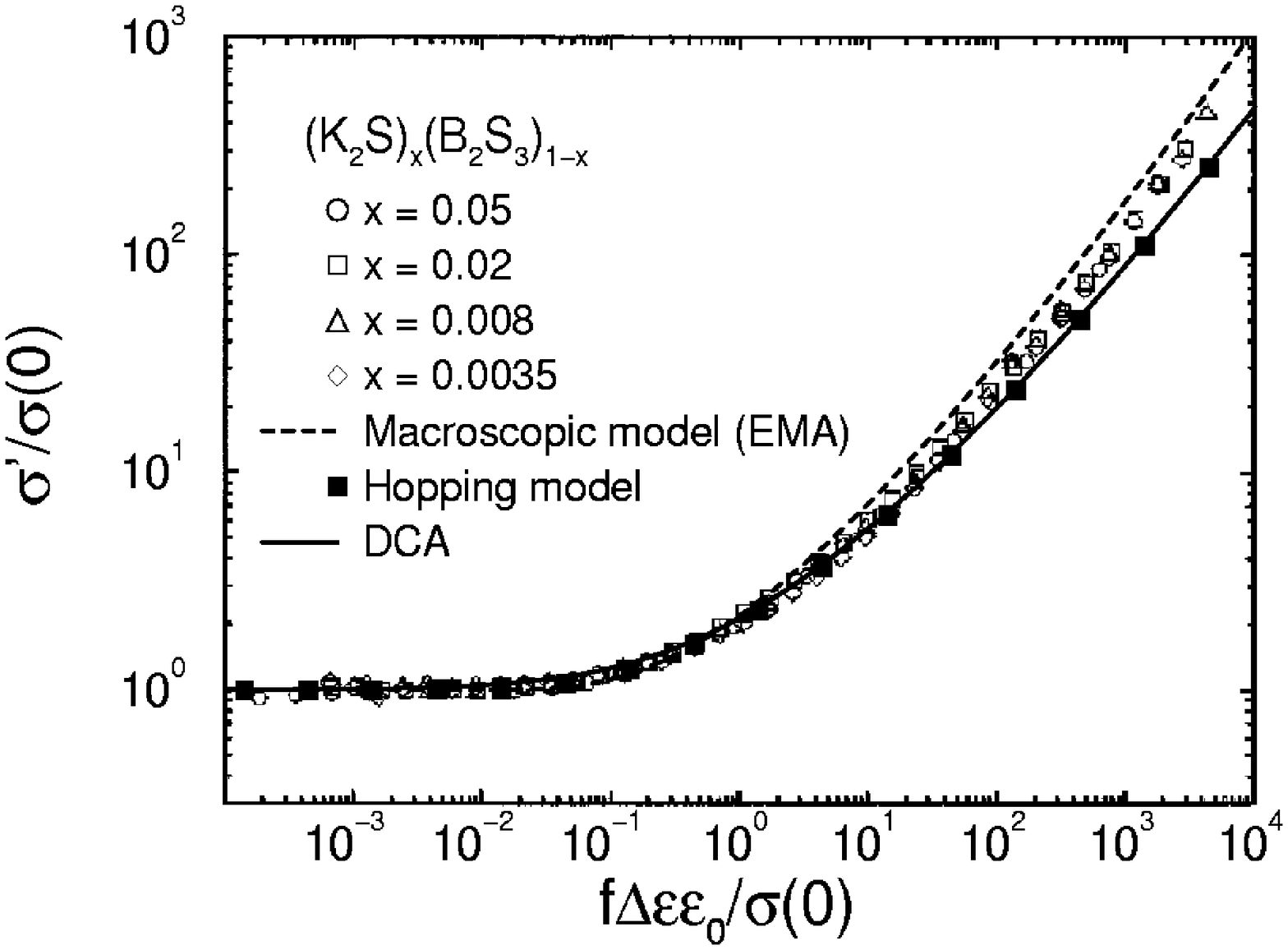}\hfil}
   \vspace{.1cm} 
\caption{Experimental data for AC conduction of 3 ionic systems scaled 
according to Eq.\ (\ref{3}) ($f=\omega/2\pi$):\\ 
a) ${\rm (Na_2O)_x(GeO_2)_{1-x}}$ (Sidebottom \protect\cite{sid99}) \\ 
b)  ${\rm 0.4Ca(NO_3)_2-0.6KNO_3}$ (Howell et. al. \protect\cite{sid99,how74}) \\ 
c)  ${\rm (K_2S)_x(B_2S_3)_{1-x}}$ (Patel \protect\cite{sid99,pat93}).\\ 
Experimental data are compared 
to the macroscopic model (represented by EMA), the universal AC conductivity
of the symmetric hopping model in the extreme disorder limit (numerical
data), and DCA (Eq.\ (\ref{5})). 
(DCA is scaled to agree with the numerical data for the hopping model, 
since it cannot  be scaled by using Eq.\ (\ref{3}) ($\de = \infty$).) 
} 
\label{fig2} 
\end{figure} 

In Fig. 2 the data discussed by Sidebottom \cite{sid99} are 
compared to the macroscopic model represented by the EMA 
universality equation (\ref{4}) (dashed line), to the hopping 
model simulations in the extreme disorder limit (filled squares), 
and to the DCA (full line, Eq.\ (\ref{5})). Figure 2a shows Sidebottom's own
data for ${\rm (Na_2O)_x(GeO_2)_{1-x}}$ glasses, Fig. 2b shows data for the
ionic melt ${\rm 0.4Ca(NO_3)_2-0.6KNO_3}$ measured by Howell and coworkers 
\cite{sid99,how74}, and Fig. 2c shows Patel's 
data for the thioborate glass system ${\rm (K_2S)_x(B_2S_3)_{1-x}}$ 
\cite{sid99,pat93}. 
In all three systems we find a frequency range where 
experimental data agree well with the symmetric hopping model. 

The symmetric hopping model is an extremely simple model which 
contains no fitting parameters but still gives a good fit to 
experimental data as shown above. 
The model does not include  Coulomb interactions and it allows an arbitrary 
number of charge carriers at each site. We note, however, that the model 
is mathematically equivalent to that obtained by linearizing 
(with respect to the electric field) a hopping model with energy 
disorder and Fermi-statistics \cite{bot85}. 
The macroscopic model does not work very well for the ionic 
systems in Fig. 2. This is interesting because the macroscopic 
model via Gauss' law includes all effects of Coulomb 
interactions. Apparently, the continuum description behind this 
model does not appropriately reflect the actual microscopic 
disorder. 
 
To summarize, we have briefly outlined the history of AC scaling 
in disordered solids and showed that the scaling version Eq.\ 
(\ref{3}) must be obeyed whenever conductivity scaling is 
possible at all (TTSP obeyed). We have presented a new analytical 
approximation to the universal AC conductivity of hopping in the 
extreme disorder limit, the diffusion cluster approximation (DCA), and 
shown that it gives an excellent fit to simulations \cite{dca}. 
Also, DCA (and thereby the extreme disorder limit of the symmetric hopping 
model) agrees well with the ionic data discussed by Sidebottom \cite{sid99},
in a frequency range depending on the system.

\acknowledgements 
This work was supported in part by the Danish Natural Science 
Research Council.


\begin{references}
\bibitem{diel_def} In solids with non-zero DC conductivity the 
dielectric constant $\epo$ is defined by $\sio-\siDC\ =\ i\omega\ 
[\epo-\epsilon_\infty]$ where $\epsilon_\infty$ is the high 
frequency dielectric constant. Writing $\epo=\epsilon'(\omega)- 
i\epsilon''(\omega)$, the quantity $\epsilon''(\omega)$ is 
referred to as the dielectric loss. The dielectric loss strength 
$\Delta\epsilon$ is the dimensionless quantity $[\epsilon(0)- 
\epsilon_\infty]/\enul$, where $\enul$ is the vacuum 
permittivity.
\bibitem{0.6} Sometimes exponent 0.6 is reported when only few 
decades of AC conductivity are observed; in practice the 
exponent depends somewhat on the frequency-range covered because 
$\sio$ is not an exact power-law.
\bibitem{anal_note} Besides TTSP the only assumption made is 
analyticity or, more precisely, the existence of a regular first 
order term in the expansion of $\sit(\omt)$. Note that, while the 
existence of this regular first order term is certain, the 
function $\sit(\omt)$ in many models is actually not analytic: 
For instance, in hopping models the low-frequency expansion looks 
like this: $\sio=\siDC+A(i\omega)+B(i\omega)^{3/2}+...$ (this 
non-analyticity expresses the existence of long time tails) 
\cite{bot85}.
\bibitem{BC} In \cite{dyr96} (where calculations were done 
for imaginary frequencies) the universal curve was found to 
be slightly closer to PPA than in figure 1b. This minor difference is due 
to different boundary conditions; we believe the present periodic boundary 
conditions give results which for any finite lattice are closer to the true 
bulk conductivity.
\bibitem{ema_units} Rationalized units are used here 
(conductivity = jump rate on a homogeneous lattice) \cite{dyr94}.
\bibitem{dca} The DCA may also be applied to the macroscopic 
model, because the reason for universality also for this model is 
percolation \cite{dyr93}. For the macroscopic model EMA leads to 
Eq.\ (\ref{4}) whenever $d_0>1$, which is the DCA-prediction for 
this model. Consequently, DCA works very well even for the 
macroscopic model.
\end{references}
\end{document}